\documentclass[12pt,a4paper]{article}
\usepackage{graphicx}
\usepackage{setspace}
\usepackage{natbib}
\usepackage{amssymb}
\usepackage{epsfig,color}
\usepackage{fancyhdr}

\begin{document}
% \title{Mass functions of clumps}
% \maketitle
% \makebox[3\width]{%
% A poster report, presented at the conference}
% \par
% \makebox[1\width]{%
% {\it Galactic Scale Star Formation: Observation meets theory,} Heidelberg, July 30 -August 3, 2012
% } \par

\pagestyle{plain}
% \setcounter{page}{1}
% \pagenumbering{arabic}

\flushright{
{\bf \emph{A poster report, presented at \\
the 2013 MPIA summer conference ``Phases of the ISM''\\
Heidelberg, July 29 -August 1, 2013}}
}
\vspace*{24pt}
\begin{center}
{\LARGE \bf Relationship between molecular cloud structure and density PDFs\vspace*{18pt}\\}
{\Large Orlin Stanchev$^{\,1}$, Sava Donkov$^{\,2}$, Todor V. Veltchev$^{\,1,\,3}$, Rahul Shetty$^{\,3}$ \vspace*{10pt}}
\flushleft{\large \it
$^1$ Faculty of Physics, University of Sofia, Bulgaria \\
$^2$ Department of Applied Physics, Technical University, Sofia, Bulgaria \\
$^3$ Institute of Theoretical Astrophysics, Heidelberg, Germany \vspace*{6pt}\\}
{\footnotesize {\bf E-mails:}~o\_stanchev@phys.uni-sofia.bg, savadd@tu-sofia.bg \vspace{12pt}\\}
\end{center}

\begin{quote}
{\bf Abstract:} Volume and column density PDFs in molecular clouds are important diagnostics for understanding their general structure. We developed a novel approach to trace the cloud structure by varying the lower PDF cut-off and exploring a suggested mass-density relationship with a power-law index $x^\prime$. The correspondence of $x^\prime$ as a function of spatial scale to the slope of the high-density PDF tail is studied. To validate the proposed model, we use results from hydrodynamical simulations of a turbulent self-gravitating cloud and recent data on dust continuum emission from the {\sc Planck} mission.
\end{quote}

\flushleft

\section{The density PDF in molecular clouds}
\begin{figure}[!ht]
\begin{center}
\includegraphics[angle=0, width=0.6\textwidth]{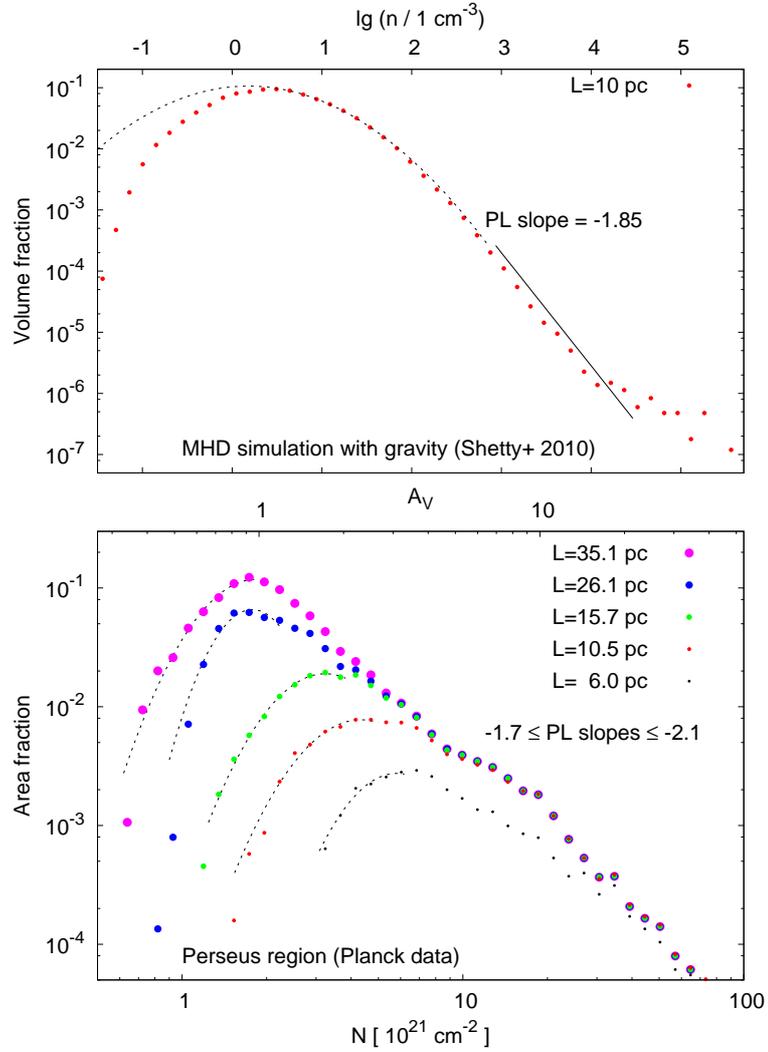}
\caption{Examples of volume (top) and column (bottom) density PDFs in regions of size $L$, consisting of lognormal part (dashed) and a power-law (PL) tail.}
\label{fig_pdfs}
\end{center}
\end{figure}

Various processes that play a key role in physics of dense, mainly molecular phase of the interstellar medium are imprinted in its velocity and density distributions. Molecular clouds (MCs) are objects of special interest since they are associated with sites of star formation. As shown in numerous studies in the last decade, the probability density function (PDF) of the gas density $\rho$ in MCs is shaped by the interplay between turbulent phenomena like power spectrum of velocity and turbulence forcing and gravitational field dominated by prestellar cores emerging in denser cloud fragments. Hence the density PDF could serve as an important diagnostic tool for understanding the general structure of MCs. Typical examples of volume density (from numerical simulations) and column density (from observations) PDFs are shown in Fig. \ref{fig_pdfs}. Their shape can be described as a combination of lognormal function and a power-law (PL) tail in the high-density regime. \vspace*{6pt}

Let $p(s)\equiv p(\ln( \rho/\rho_0))$ be the volume-weighted density PDF (of arbitrary shape) in a MC of mass $M_{\rm c}$, volume $V_{\rm c}$ and mean density $\rho_0=M_{\rm c}/V_{\rm c}$. We develop a simple approach to trace the cloud structure in two steps: 
\begin{itemize}
 \item Varying a lower cut-off $s^\prime$ of the density PDF. This step mimics observational mapping through varying the {\it column-density} threshold. 
 \item Exploring a suggested relationship between mass $M^\prime$ and mean density $\langle\rho\rangle^\prime$ derived over the cut-off $s^\prime$ or between mass $m$ and mean density $\rho$ of typical objects at this spatial scale. Such relationships are implied by the established mass-size relationship in MCs \citep[e.g.,][]{LAL_10, BP_ea_12} and the density scaling law of clouds and cloud fragments of \citet{Larson_81}.
\end{itemize}

\section{Modeling the cloud structure}
\subsection{Global structure}
The mass, the volume and the effective size of a (not necessarily connected) subregion are determined by a log-density cut-off $s^\prime$:
\begin{equation}
\label{Mpr_Vpr_Lpr}
 M^\prime=M_{\rm c} \int_{s^\prime}^\infty \exp(s)p(s)\,ds~,~~V^\prime=V_{\rm c} \int_{s^\prime}^\infty p(s)\,ds~,~~L^\prime=(V^\prime)^{1/3}~,
\end{equation} 
where $L^\prime$ is interpreted as a spatial scale. Then the mean density of this scale and its relation to its mass and density are given by:
\begin{equation}
\label{rhopr}
 \langle\rho\rangle^\prime=\frac{M^\prime}{V^\prime}=\rho_0\frac{\int_{s^\prime}^\infty\exp(s)p(s)\,ds}{\int_{s^\prime}^\infty p(s)\,ds}~,
\end{equation}

\begin{equation}
\label{M_rho_V-norm}
\frac{M^\prime}{M_{\rm c}}=\frac{\langle\rho\rangle^\prime}{\rho_0}\,\frac{V^\prime}{V_{\rm c}}
\end{equation}
Our {\it basic assumption} in this approach is that a power-law mass-density relationship holds at each scale:
\begin{equation}
\label{Basic_assump}
 \langle\rho\rangle^\prime/\rho_0=\big(M^\prime/M_{\rm c}\big)^{x^\prime}~,
\end{equation}
where $x^\prime=x^\prime(s^\prime)$ is labeled ``structure index'' since it traces the global MC structure.
 
\subsection{Local structure}
Statistical clumps with mass $m$, density $\rho$ and volume $V$ are considered at scale $L^\prime$ with PDF $p^\prime(s)$ and defined by the relations:
\begin{equation}
\label{local_str}
 \frac{m}{m_0^\prime}=\frac{\rho}{\rho_0^\prime}\,\frac{V}{V_0^\prime}~,~~~\frac{\rho}{\rho_0^\prime}=\Bigg(\frac{m}{m_0^\prime}\Bigg)^x~,
 \end{equation}
where the relations between normalization units $m_0^\prime$, $\rho_0^\prime\equiv\langle\rho\rangle^\prime$ and $V_0^\prime$ are calculated from the conservation laws of mass and volume and $x\equiv x^\prime$ is assumed. \vspace*{9pt}\\ 

In this approach, we {\it postulate an invariant} typical for compressible turbulent medium with scaling laws of velocity $u_l\propto l^\beta$ and density $\rho_l\propto l^\alpha$, for statistical clumps:
\begin{equation}
\label{rho_u_l-inv}
\rho_l\,u_l^\delta/l={\rm const}(l)~,  
\end{equation}
where $l=V^{1/3}$, $\alpha+\delta\beta-1=0$ (from dimensional analysis) and $\alpha=3x/(1-x)$ (assuming self-similarity within the considered scale; see \citealp{DVK_11}).

\subsection{Natural relations between structure index $x^\prime$, density PDF, cut-off $s^\prime$ and mean cloud density $\rho_{0}$.}

From the expressions of fractional mass and volume of structure, delineated by cut-off $s^\prime$, and Eq.~\ref{Mpr_Vpr_Lpr} one gets: 
$$\frac{M^\prime}{M_{c}}= \int_{s^\prime}^\infty \exp(s)p(s)\,ds~~\Longrightarrow~~\frac{d}{ds^\prime}\frac{M^\prime}{M_{c}}= -\exp(s^\prime)p(s^\prime)~,$$

$$\frac{V^\prime}{V_{c}}= \int_{s^\prime}^\infty p(s)\,ds~~\Longrightarrow~~\frac{d}{ds^\prime}\frac{V^\prime}{V_{c}}= -p(s^\prime)$$

$$\Longrightarrow~~ d\Bigg(\frac{M^\prime}{M_{c}}\Bigg)=\exp(s^\prime)d\Bigg(\frac{V^\prime}{V_{c}}\Bigg)~;$$

and then from Eq.~\ref{M_rho_V-norm}:

$$\frac{M^\prime}{M_{\rm c}}=\frac{\langle\rho\rangle^\prime}{\rho_0}\,\frac{V^\prime}{V_{\rm c}}~~\Longrightarrow~~\frac{V^\prime}{V_{c}}=\frac{M^\prime}{M_{c}}\,\Bigg(\frac{\langle\rho\rangle^\prime}{\rho_0}\Bigg)^{-1}$$

$$\Longrightarrow~~ d\Bigg(\frac{V^\prime}{V_{c}}\Bigg)= \Bigg(\frac{\langle\rho\rangle^\prime}{\rho_0}\Bigg)^{-1}\,d\Bigg(\frac{M^\prime}{M_{c}}\Bigg)-\Bigg(\frac{\langle\rho\rangle^\prime}{\rho_0}\Bigg)^{-2}\,\frac{M^\prime}{M_{\rm c}}\,d\Bigg(\frac{\langle\rho\rangle^\prime}{\rho_0}\Bigg)~.$$

Combining the last two differential relations, we obtain:

$$d\Bigg(\frac{M^\prime}{M_{c}}\Bigg)= \Bigg(\frac{\langle\rho\rangle^\prime}{\rho_0}\Bigg)^{-1}\,\exp(s^\prime)\,d\Bigg(\frac{M^\prime}{M_{c}}\Bigg)-\Bigg(\frac{\langle\rho\rangle^\prime}{\rho_0}\Bigg)^{-2}\,\frac{M^\prime}{M_{\rm c}}\,\exp(s^\prime)\,d\Bigg(\frac{\langle\rho\rangle^\prime}{\rho_0}\Bigg)~.$$

Finally, after simple algebraic manipulation, we derive the equation:

\begin{equation}
\label{lnrho_lnM-relation}
d\ln\Bigg(\frac{\langle\rho\rangle^\prime}{\rho_0}\Bigg)=\Bigg[1-\exp(-s^\prime)\frac{\langle\rho\rangle^\prime}{\rho_0}\Bigg]\,d\ln\Bigg(\frac{M^\prime}{M_{c}}\Bigg)~.
\end{equation}

On the other hand, from Eq.~\ref{M_rho_V-norm}:

$$\ln\Bigg(\frac{\langle\rho\rangle^\prime}{\rho_0}\Bigg)= x^\prime\,\ln\Bigg(\frac{M^\prime}{M_{c}}\Bigg)$$

$$\Longrightarrow~~d\ln\Bigg(\frac{\langle\rho\rangle^\prime}{\rho_0}\Bigg)= x^\prime\,d\ln\Bigg(\frac{M^\prime}{M_{c}}\Bigg)+\ln\Bigg(\frac{M^\prime}{M_{c}}\Bigg)\,dx^\prime~,$$

and because

$$d\ln\Bigg(\frac{M^\prime}{M_{c}}\Bigg)=\Bigg[1-\exp(-s^\prime)\frac{\langle\rho\rangle^\prime}{\rho_0}\Bigg]^{-1}\,d\ln\Bigg(\frac{\langle\rho\rangle^\prime}{\rho_0}\Bigg)~~and~~\ln\Bigg(\frac{M^\prime}{M_{c}}\Bigg)=\frac{1}{x^\prime}\,\ln\Bigg(\frac{\langle\rho\rangle^\prime}{\rho_0}\Bigg)~,$$

it follows that:

\begin{equation}
\label{full_xpr_eq}
\Bigg[1-\exp(-s^\prime)\frac{\langle\rho\rangle^\prime}{\rho_0}-x^\prime\Bigg]\,d\ln\Bigg(\frac{\langle\rho\rangle^\prime}{\rho_0}\Bigg)=\ln\Bigg(\frac{\langle\rho\rangle^\prime}{\rho_0}\Bigg)\,\Bigg[1-\exp(-s^\prime)\frac{\langle\rho\rangle^\prime}{\rho_0}\Bigg]\,d\ln(x^\prime)~.
\end{equation}

This is the full equation connecting structure index $x^\prime$, density PDF, log-density cut-off $s^\prime$ and mean cloud density $\rho_{0}$. We point out that the volume weighted PDF is of most general shape, with the only restriction that $p(s)$ must be a continuous function. \vspace*{6pt}
 
Equation \ref{full_xpr_eq} is too complicated to be solved directly, but if the function $x^\prime=x^\prime(s^\prime)$ changes negligibly in a range $[s_{0},\infty]$, one can neglect the term $\ln(M^\prime/M_{\rm c})\,dx^\prime$ and thus $d\ln(\langle\rho\rangle^\prime/\rho_{0})\approx x^\prime\,d\ln(M^\prime/M_{\rm c})$. Recalling Eq.~\ref{lnrho_lnM-relation}, we obtain a simple formula:

\begin{equation}
\label{approx_xpr_eq}
x^\prime\approx 1-\exp(-s^\prime)\frac{\langle\rho\rangle^\prime}{\rho_0}= 1-\exp(-s^\prime)\,\rho_0\frac{\int_{s^\prime}^\infty\exp(s)p(s)\,ds}{\int_{s^\prime}^\infty p(s)\,ds}~.
\end{equation}

At last, it is good to note, that the structure index $x^\prime$ can be derived at first hand from Eq. \ref{Basic_assump} by taking logarithm and replacing $M^\prime/M_{c}$ and $\langle\rho\rangle^\prime/\rho_{0}$ from Eqs. \ref{Mpr_Vpr_Lpr} and \ref{rhopr}. We will have:

\begin{equation}
\label{direct_xpr_eq}
x^\prime= 1-\frac{\ln\bigg(\int_{s^\prime}^\infty p(s)\,ds\bigg)}{\ln\bigg(\int_{s^\prime}^\infty\exp(s)p(s)\,ds\bigg)}~.
\end{equation}

Indeed, this is the exact solution of Eq.~\ref{full_xpr_eq}, while Eq.~\ref{approx_xpr_eq} gives the solution when $dx^\prime/ds^\prime=0$, e.g. in the range of a power-law tail of the PDF.

\section{Interpretation of the clump mass-density diagram}
\begin{figure}[!ht]
\begin{center}
\includegraphics[angle=0, width=0.55\textwidth]{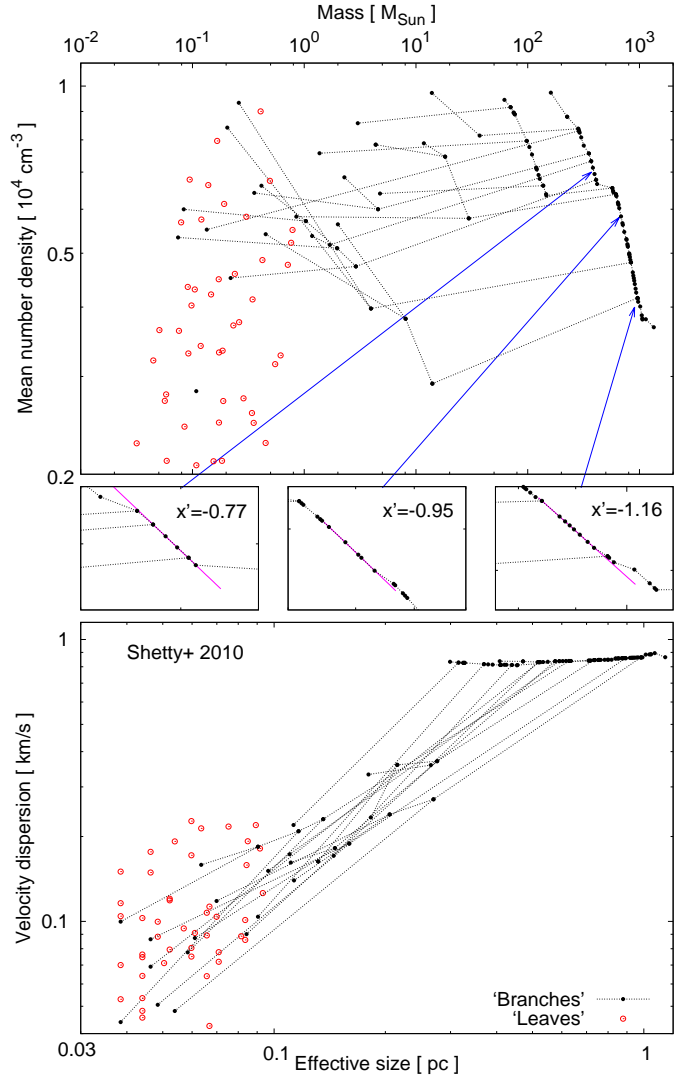}
\caption{Mass-density (top) and velocity-size (bottom) diagram of MC clumps, delineated by use of dendrogram technique.}
\label{fig_scalings}
\end{center}
\end{figure}

In terms of the dendrogram technique of clump extraction \citep{Rosolowsky_ea_08}, we present the MC structure as a `tree' of spatially separated or embedded substructures (`branches') and single small objects with no or negligible substructure (`leaves'). The inclination of a `branch' in respect to the abscissa on the $\langle\rho\rangle^\prime-M^\prime$ diagram is given by $x^\prime$ (Fig. \ref{fig_scalings}, top and middle). Two types of substructures are identified both from statistical and morphological consideration: {\it global} and {\it local}.  

\begin{itemize}
 \item[$\bullet$] {\it Local substructure:} In morphological treatment, it is recognized as a diverging point $(\langle\rho\rangle^\prime,\,M^\prime)$ of the `tree' which produces - typically, - a main `branch' containing most of the volume and mass and a `leaf' with density $\rho<\langle\rho\rangle^\prime$ and mass $m$ about 2 orders of magnitude lower than $M^\prime$. The statistical interpretation of the `leaf' is the most typical clump generated at the local scale $L^\prime$ while the main `branch' represents the rest of the latter.
 \item[$\bullet$] {\it Global substructure:} Morphologically, it is traced by series of main `branches' (which only `leaves' deviate from). From statistical perspective, it is represented by a series of spatial scales with decreasing size as defined through successive PDF cut-offs $s^\prime$. Note that the mean densities even at largest scales correspond to the tail of the PDF (cf. Figs. \ref{fig_pdfs} /top/ and \ref{fig_scalings} /top/).
\end{itemize}

\section{Discussion}

 \begin{itemize}
 \item Every `branch' with significant substructure in the series of decreasing masses is characterized by a decreasing $|x^\prime|$ (Fig. \ref{fig_scalings}, middle) which corresponds to a steeper PL tail of the PDF. We have obtained from modeling of the global structure:
 \[  x^\prime=1-\exp(-s^\prime)\,\langle\rho\rangle^\prime/\rho_0~,       \]
 which yields $x^\prime=1/(q+1)$ for the density regime of PL tail with slope $q$. Note that $q=-1.85$ from the considered simulations (Shetty et al. 2010) and thus $x^\prime\approx-1.18$, in excellent agreement with the results for the largest `branches' in the dendrogram. Similar slopes of the PDF tail are theoretically predicted and confirmed in numerical simulations by Girichidis et al. (2013). 
 \item Adopting $x\equiv x^\prime=-1.18$ in describing the local structure, one gets for the density scaling index $\alpha=-1.62$. Assuming turbulent invariants with $\delta=3$ or $\delta=2$, we obtain for the velocity scaling index $\beta=0.87$ or $\beta=1.31$, respectively. These results are consistent with the clump velocity-size diagram (Fig.~\ref{fig_scalings}, bottom) from the used simulations of \citet{Shetty_ea_10}.
 \item It can be demonstrated that the spatial scales delineated by varying the density cut-off $s^\prime$ form a fractal hierarchy \citep[cf.][]{Elme_97}. Therefore their mass function, interpreted as mass distribution of embedded (not necessarily connected) cloud fragments, has a typical slope $\Gamma=-1$. If one considers a time-weighted mass function, adopting the free-fall time as a typical evolutionary time-scale, then the relation $\Gamma=-1+x^\prime/2$ can be derived \citep{DSV_12}. This is an important result since it links evolution of the cloud substructure with the physical parameters of density statistics. 
 \item The PDFs derived in embedded substructures of Perseus region demonstrate, in principle, the same morphological and statistical behavior. The most significant result is that the PDF in each subregion retains the same shape which is a combination of lognormal function and PL tail.
 \end{itemize}

{\it Acknowledgement:} T.V. acknowledges support by the {\em Deutsche Forschungsgemeinschaft} (DFG) under grant KL 1358/15-1.


\newpage
\begin{thebibliography}{}

\bibitem[\protect\citeauthoryear{Ballesteros-Paredes et al.}{2012}]{BP_ea_12}
Ballesteros-Paredes, J., D'Alessio, P., Hartmann, L., 2012, MNRAS, 427, 2562
\bibitem[\protect\citeauthoryear{Donkov, Stanchev \& Veltchev}{2012}]{DSV_12}	
Donkov, S., Stanchev, O., Veltchev, T., 2012, Proc. of the VIII Serbian-Bulgarian Astron. Conf., Leskovac, Serbia, May 8-12, 2012, eds. M. K. Tsvetkov, M. S. Dimitrijevic, K. Tsvetkova, O. Kounchev, Z. Mijajlovic (arXiv 1206.1444)
\bibitem[\protect\citeauthoryear{Donkov, Veltchev \& Klessen}{2011}]{DVK_11}
Donkov, S., Veltchev, T., Klessen, R. S., 2011, MNRAS, 418, 916
\bibitem[\protect\citeauthoryear{Elmegreen}{1997}]{Elme_97}
Elmegreen, B. G., 1997, ApJ, 486, 944
\bibitem[\protect\citeauthoryear{Girichidis et al.}{2013}]{Girichidis_ea_13}
Girichidis, P., Konstandin, L., Klessen, R., Whitworth, A., 2013, ApJ (in preparation)
\bibitem[\protect\citeauthoryear{Larson}{1981}]{Larson_81}
Larson, R., 1981, MNRAS, 194, 809
\bibitem[\protect\citeauthoryear{Lombardi, Alves \& Lada}{2010}]{LAL_10}
Lombardi, M., Alves, J., Lada, C., 2010, A\&A, 519, 7 
\bibitem[\protect\citeauthoryear{Rosolowsky et al.}{2008}]{Rosolowsky_ea_08}
Rosolowsky, E., Pineda, J., Kauffmann, J., Goodman, A., 2008, ApJ, 679, 1338
\bibitem[\protect\citeauthoryear{Shetty et al.}{2010}]{Shetty_ea_10}
Shetty, R., Collins, D., Kauffmann, J., Goodman, A., Rosolowsky, E., Norman, M., 2010, ApJ, 712, 1049
  

\end{thebibliography}
\end{document}